\documentclass[10pt,a4papper]{article}
\usepackage[utf8]{inputenc}
\usepackage{indentfirst}
\usepackage[]{graphicx}
\usepackage{geometry} 
\begin{document}
	\title{\textbf{Calculation of the width of the decay $\tau \to K^{-} K^{0} \nu_{\tau}$ in the extended NJL model with estimation of the contribution of the kaon final state interaction}}
	\author{M. K. Volkov\footnote{volkov@theor.jinr.ru}, A. A. Pivovarov\footnote{tex$\_$k@mail.ru}\\
		\small
		\emph{BLTP, Joint Institute for Nuclear Research, Dubna, 141980, Russia}}
	\date{}
	\maketitle
	\small

\begin{abstract}
    The branching fraction of the decay $\tau \to K^{-} K^{0} \nu_{\tau}$ is calculated in the framework of the extended Nambu--Jona-Lasinio model. The contact and vector channels are considered. The contributions to this process of the $\rho$ meson in the ground and first radially excited states are taken into account in the vector channel. The obtained results are in satisfactory agreement with the experimental data. Taking into account the kaon interaction in the final state results in insignificant corrections, which are not beyond the scope of the model uncertainties.
\end{abstract}

\large
\section{Introduction}
    In the recent works considering tau lepton decays into two pseudoscalar mesons \cite{Volkov:2020dvz,Volkov:2021bfx,Volkov:2021sma}, it was shown that in the description of these processes in the framework of the Nambu--Jona-Lasinio model (NJL) \cite{Volkov:1986zb,Volkov:2005kw} the meson interaction in the final state plays an important role. Especially, this concerns the processes with the produced light pions. Then the correction to the decay widths can reach 30\%. This correction decreases with an increase in the total mass of the produced mesons. For example, in the case of the kaon and $\eta$ meson production the correction becomes about 15\%. However, in the case of two kaons in the final state, it turns out to be about 6\% and becomes negligible in the framework of the model uncertainties, which can be estimated at the level of 15\%. Thus, one can see a tendency towards a decrease in the contribution of the final state interaction with an increase in the total mass of the produced mesons and the appearance of the necessity to take into account the intermediate excited mesons.
    
    The decay considered in the present work is still attracting the attention of experimental collaborations and theoretical groups. For example, it was investigated in the recent works of the Belle and BaBar collaborations \cite{Belle:2014mfl,BaBar:2018qry}. Also, this decay was considered in numerous theoretical works \cite{Li:1996md,Dubnicka:2010grh,Dai:2018thd,Gonzalez-Solis:2019lze} using Resonance Chiral Perturbation Theory, Vector Meson Dominance model, angular momentum algebra, etc. 

\section{The Lagrangian of the extended NJL model}
    A fragment of the quark-meson Lagrangian of the extended NJL model with the necessary vertices takes the form~\cite{Volkov:2005kw,Volkov:1996br,Volkov:1996fk,Volkov:2017arr}:
    \begin{eqnarray}
    	\Delta L_{int} & = &
    	\bar{q} \left[ \frac{1}{2} \gamma^{\mu} \sum_{j = \pm}\lambda_{j}^{\rho} \left(A_{\rho}\rho^{j}_{\mu} + B_{\rho}\rho^{'j}_{\mu}\right) + i \gamma^{5} \sum_{j = \pm,0} \lambda_{j}^{K} \left(A_{K}K^{j} + B_{K}K'^{j}\right) \right]q,
    	\end{eqnarray}
    where $q$ and $\bar{q}$ are the u, d and s quark fields with the constituent masses $m_{u} \approx m_{d} = 280$~MeV, $m_{s} = 420$~MeV. The excited meson states are marked with a prime. The factors $A$ and $B$ take the form:
    \begin{eqnarray}
    \label{verteces1}
    	A_{M} & = & \frac{1}{\sin(2\theta_{M}^{0})}\left[g_{M}\sin(\theta_{M} + \theta_{M}^{0}) +
    	g_{M}^{'}f_{M}(k_{\perp}^{2})\sin(\theta_{M} - \theta_{M}^{0})\right], \nonumber\\
    	B_{M} & = & \frac{-1}{\sin(2\theta_{M}^{0})}\left[g_{M}\cos(\theta_{M} + \theta_{M}^{0}) +
    	g_{M}^{'}f_{M}(k_{\perp}^{2})\cos(\theta_{M} - \theta_{M}^{0})\right].
    \end{eqnarray}
    The index $M$ designates the appropriate meson.
    
    The form factor describing the first radially excited meson states takes the form~\cite{Volkov:2017arr}:
    \begin{eqnarray}
        f\left(k_{\perp}^{2}\right) = \left(1 + d {\bf k}^{2}\right)\Theta(\Lambda^{2} - {\bf k}^2),
    \end{eqnarray}
    where $d$ is the slope parameter depending on the quark composition of the appropriate meson \cite{Volkov:2017arr} and $k$ is the relative momentum of the quarks in a meson.
    
    The parameters $\theta_{M}$ are the mixing angles appearing as a result of the diagonalization of the Lagrangian with the mesons in the ground and first radially excited states \cite{Volkov:2017arr}:
    \begin{eqnarray}
    \label{angels}
    	&\theta_{\rho} = 81.8^{\circ}, \quad \theta_{\rho}^{0} = 61.5^{\circ}, \quad \theta_{K} = 58.11^{\circ}, \quad \theta_{K}^{0} = 55.52^{\circ}.&
    \end{eqnarray}
    
    The matrices $\lambda$ are the linear combinations of the Gell-Mann matrices.
    
    The coupling constants are:
    \begin{eqnarray}
    	\label{Couplings}
    	g_{\rho} =  \left(\frac{3}{2I_{20}}\right)^{1/2}, &\quad& g_{\rho}^{'} =  \left(\frac{3}{2I_{20}^{f^{2}}}\right)^{1/2}, \nonumber\\
    	g_{K} =  \left(\frac{Z_{K}}{4I_{11}}\right)^{1/2}, &\quad& g_{K}^{'} =  \left(\frac{1}{4I_{11}^{f^{2}}}\right)^{1/2},
    \end{eqnarray}
    where $Z_{K}$ is the additional renormalization constant appearing in the $K-K_{1}$ transitions:
    \begin{eqnarray}
    \label{Zk}
    	Z_{K} & = & \left(1 - \frac{3}{2} \frac{(m_{u} + m_{s})^{2}}{M_{K_{1A}}^{2}}\right)^{-1}, \nonumber\\
    	M_{K_{1A}} & = & \left(\frac{\sin^{2}{\alpha}}{M^{2}_{K_{1}(1270)}} + \frac{\cos^{2}{\alpha}}{M^{2}_{K_{1}(1400)}}\right)^{-1/2}.
    \end{eqnarray}
    The split of the state $K_{1A}$ into two physical mesons $K_{1}(1270)$ and $K_{1}(1400)$ with the mixing angle $\alpha = 57^{\circ}$ \cite{Volkov:2019yhy} is taken into account here. The meson masses used there are $M_{K_{1}(1270)} = 1253 \pm 7$~MeV and $M_{K_{1}(1400)} = 1403 \pm 7$~MeV~\cite{Zyla:2020zbs}.
    
    The integrals appearing in the quark loops as a result of the renormalization of the Lagrangian are:
    \begin{eqnarray}
    	I_{n_{1}n_{2}}^{f^{m}} =
    	-i\frac{N_{c}}{(2\pi)^{4}}\int\frac{f^{m}({\bf k}^{2})}{(m_{u}^{2} - k^2)^{n_{1}}(m_{s}^{2} - k^2)^{n_{2}}}\Theta(\Lambda^{2} - {\bf k}^2)
    	\mathrm{d}^{4}k,
    \end{eqnarray}
    where $\Lambda = 1.03$~GeV is the three-dimensional cutoff parameter of the quark loops~\cite{Volkov:2017arr}.
    
\section{The process $\tau \to K^{-} K^{0} \nu_{\tau}$ in the extended NJL model}
\begin{figure}[h]
    	\center{\includegraphics[scale = 0.8]{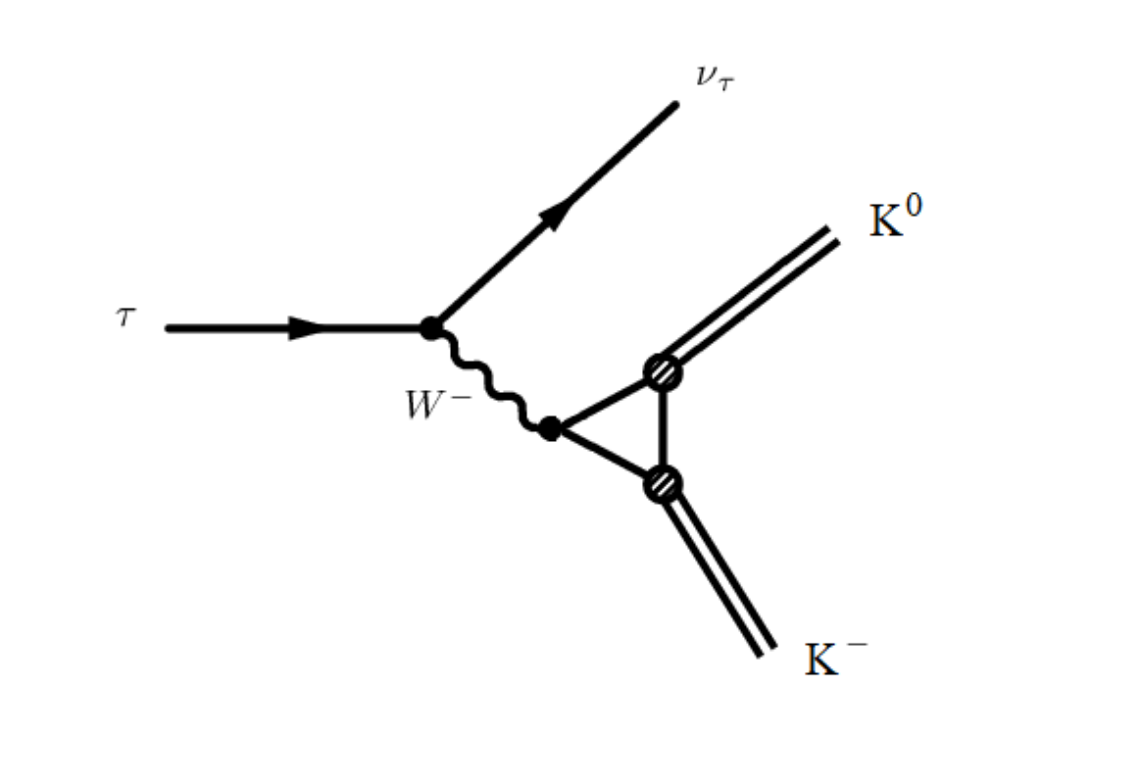}}
    	\caption{The contact diagram.}
    	\label{Contact}
    \end{figure}

\begin{figure}[h]
    	\center{\includegraphics[scale = 0.9]{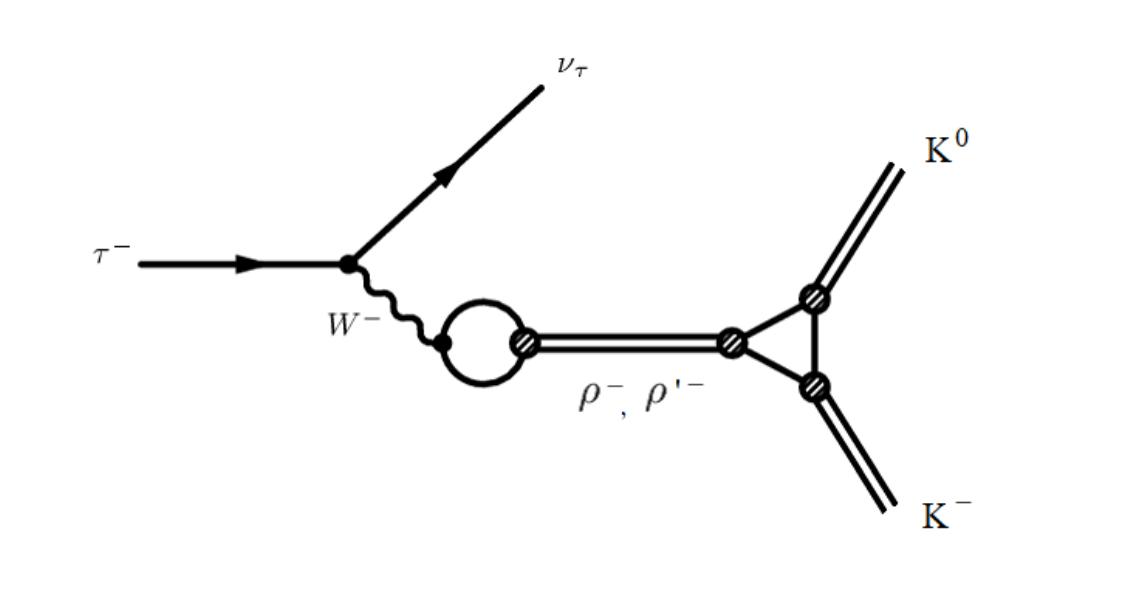}}
    	\caption{The diagram with the intermediate mesons.}
    	\label{Interm}
    \end{figure}

    The diagrams describing the process $\tau \to K^{-} K^{0} \nu_{\tau}$ are shown in Figs. \ref{Contact}, \ref{Interm}.

    In the NJL model, these diagrams are described by the following amplitude:
    \begin{eqnarray}
        M_{tree} & = &-2 \sqrt{2} G_{f} V_{ud} I_{11}^{KK} \left[T_{K}^{(c)} + \frac{C_{\rho}}{g_{\rho}} \frac{I_{11}^{KK\rho}}{I_{11}^{KK}} T_{K}^{(\rho)} \frac{q^2}{M_{\rho}^{2} - q^{2} - i\sqrt{q^{2}}\Gamma_{\rho}} \right. \nonumber\\
        && \left.+ \frac{C_{\rho'}}{g_{\rho}} \frac{I_{11}^{KK\rho'}}{I_{11}^{KK}} T_{K}^{(\rho')} \frac{q^2}{M_{\rho'}^{2} - q^{2} - i\sqrt{q^{2}}\Gamma_{\rho'}}\right] L_{\mu} \left(p_{K^{0}} -p_{K^{-}}\right)^{\mu},
    \end{eqnarray}
    where $G_{f}$ is the Fermi constant, $V_{ud}$ is the element of the Cabibbo–Kobayashi–Maskawa matrix, $L_{\mu}$ is the lepton current, $M_{\rho} = 775.11 \pm 0.34$~MeV and $\Gamma_{\rho} = 149.1 \pm 0.8$~MeV are the mass and full width of the meson $\rho(770)$, $M_{\rho'} = 1465 \pm 25$~MeV and $\Gamma_{\rho} = 400 \pm 60$~MeV are the mass and full width of the meson $\rho(1450)$~\cite{Zyla:2020zbs}.
    
    The constants describing the $K_{1}-K$ transitions are:
    \begin{eqnarray}
         T_{K}^{(c)} & = & 1 - \frac{\left(I_{11}^{K_{1}K}\right)^2}{I_{11}^{KK}} \frac{\left(m_{s} + m_{u}\right)^{2}}{M_{K_{1A}}^{2}}, \nonumber\\
         T_{K}^{(\rho)} & = & 1 - \frac{I_{11}^{K_{1}K\rho}I_{11}^{K_{1}K}}{I_{11}^{KK\rho}} \frac{\left(m_{s} + m_{u}\right)^{2}}{M_{K_{1A}}^{2}}, \nonumber\\
         T_{K}^{(\rho')} & = & 1 - \frac{I_{11}^{K_{1}K\rho'}I_{11}^{K_{1}K}}{I_{11}^{KK\rho'}} \frac{\left(m_{s} + m_{u}\right)^{2}}{M_{K_{1A}}^{2}}.
    \end{eqnarray}
    
    The integrals with the vertices from the Lagrangian in the numerator are:
    \begin{eqnarray}
    	I_{n_{1} n_{2}}^{M, \dots, M^{'}, \dots} =
    	-i\frac{N_{c}}{(2\pi)^{4}}\int\frac{A_{M} \dots B_{M} \dots}{(m_{u}^{2} - k^2)^{n_{1}}(m_{s}^{2} - k^2)^{n_{2}}} \Theta(\Lambda^{2} - {\bf k}^2)
    	\mathrm{d}^{4}k,
    \end{eqnarray}
    where $A_{M}, B_{M}$ are defined in (\ref{verteces1}).
    
    The constants 
    \begin{eqnarray}
        C_{\rho} & = & \frac{1}{\sin\left(2\theta_{\rho}^{0}\right)} \left[\sin\left(\theta_{\rho} + \theta_{\rho}^{0}\right) + R_{\rho}\sin\left(\theta_{\rho} - \theta_{\rho}^{0}\right)\right], \nonumber\\
        C_{\rho'} & = & \frac{-1}{\sin\left(2\theta_{\rho}^{0}\right)} \left[\cos\left(\theta_{\rho} + \theta_{\rho}^{0}\right) + R_{\rho}\cos\left(\theta_{\rho} - \theta_{\rho}^{0}\right)\right]
    \end{eqnarray}
    appear in the transitions between the W boson and the intermediate vector meson. Here $\theta$ and $\theta^{0}$ are the mixing angles of the ground and excited states. They were defined in (\ref{angels}). The values $R$ take the following form:
    \begin{eqnarray}
        R_{\rho} = \frac{I_{20}^{f}}{\sqrt{I_{20}I_{20}^{f^{2}}}}.
    \end{eqnarray}
    
    The result for the partial decay width of the considered process is:
    \begin{eqnarray}
    	Br(\tau \to K^{-} K^{0} \nu_{\tau}) & = & 13.95  \times 10^{-4}.
    \end{eqnarray}
    
    The experimental value is \cite{Zyla:2020zbs}:
    \begin{eqnarray}
    	Br(\tau \to K^{-} K^{0} \nu_{\tau})_{exp} & = & (14.86 \pm 0.34) \times 10^{-4}.
    \end{eqnarray}
    
    As one can see, the result is beyond the experimental errors. However, it is in the framework of the model uncertainties.
    
    The channel with the intermediate meson $\rho(1450)$ gives a big contribution to the process. This contribution is sensitive to the decay width of the meson ($\Gamma_{\rho} = 400 \pm 60$~MeV). Taking the value 380 MeV for this width, we obtain the result coinciding with the experimental data.
    
\section{Taking into account the interaction in the final state}
\begin{figure}[h]
    	\center{\includegraphics[scale = 0.8]{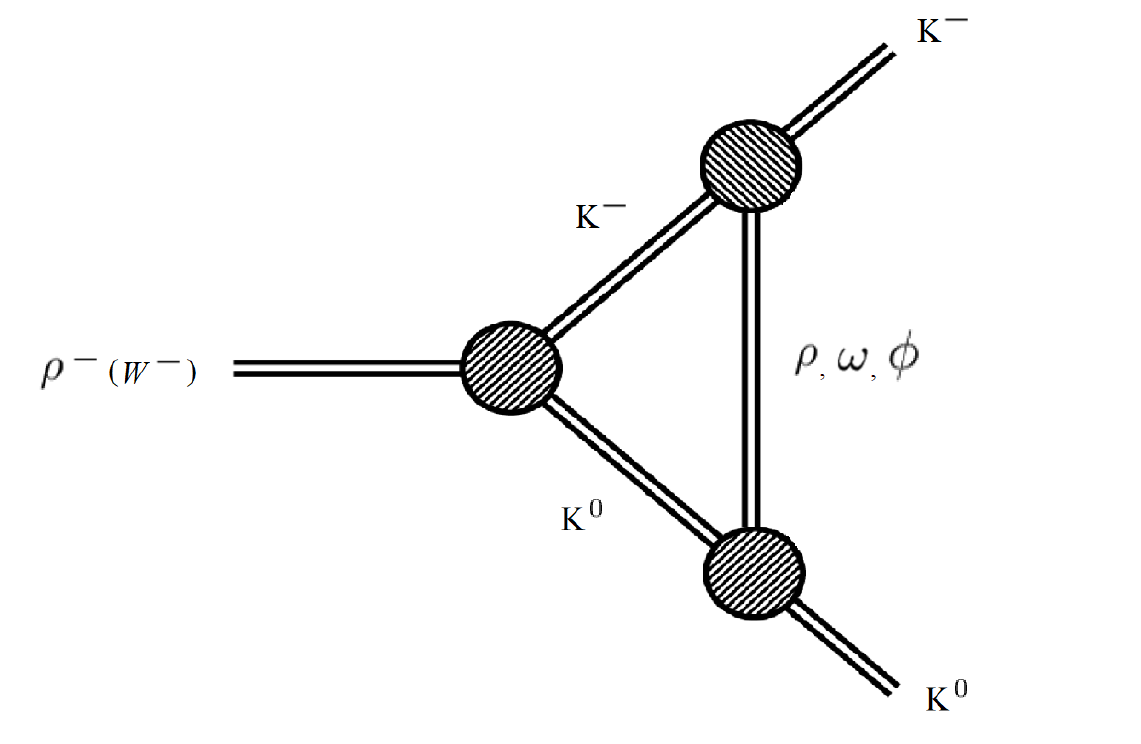}}
    	\caption{The meson triangles.}
    	\label{Triangle}
    \end{figure}

    To  take into account the meson interaction in the final state, it is necessary to consider the triangle meson diagrams with the exchange of the neutral mesons $\rho$, $\omega$ and $\phi$, as shown in Fig. \ref{Triangle}. The appropriate vertices could be found in the paper \cite{Volkov:1986zb}.
    
    The following integrals correspond to these meson triangles:
    \begin{eqnarray}
    	&F_{\mu}^{(\rho)} = \int \frac{\left(k - 2p_{K^{-}}\right)_{\lambda}\left(k + 2p_{K^{0}}\right)_{\nu}
    	\left(2k + p_{K^{0}} - p_{K^{-}}\right)_{\mu} \left(g^{\nu\lambda} 
    	- \frac{k^{\nu}k^{\lambda}}{M_{\rho}^{2}}\right)}{\left[k^{2} - M_{\rho}^{2}\right]
    	\left[(k + p_{K^{0}})^{2} - M_{K}^{2}\right]\left[(k - p_{K^{-}})^{2} - M_{K}^{2}\right]} \frac{d^{4}k}{(2\pi)^{4}},& \nonumber\\
    	&F_{\mu}^{(\omega)} = \int \frac{\left(k - 2p_{K^{-}}\right)_{\lambda}\left(k + 2p_{K^{0}}\right)_{\nu}
    	\left(2k + p_{K^{0}} - p_{K^{-}}\right)_{\mu} \left(g^{\nu\lambda} 
    	- \frac{k^{\nu}k^{\lambda}}{M_{\omega}^{2}}\right)}{\left[k^{2} - M_{\omega}^{2}\right]
    	\left[(k + p_{K^{0}})^{2} - M_{K}^{2}\right]\left[(k - p_{K^{-}})^{2} - M_{K}^{2}\right]} \frac{d^{4}k}{(2\pi)^{4}},& \nonumber\\
    	&F_{\mu}^{(\phi)} = \int \frac{\left(k - 2p_{K^{-}}\right)_{\lambda}\left(k + 2p_{K^{0}}\right)_{\nu}
    	\left(2k + p_{K^{0}} - p_{K^{-}}\right)_{\mu} \left(g^{\nu\lambda} 
    	- \frac{k^{\nu}k^{\lambda}}{M_{\phi}^{2}}\right)}{\left[k^{2} - M_{\phi}^{2}\right]
    	\left[(k + p_{K^{0}})^{2} - M_{K}^{2}\right]\left[(k - p_{K^{-}})^{2} - M_{K}^{2}\right]} \frac{d^{4}k}{(2\pi)^{4}}.&
    \end{eqnarray}
    
    As one can see, the structure of these integrals is similar to the integral obtained in the work \cite{Volkov:2020dvz}. They take the form:
    \begin{eqnarray}
    	F_{\mu}^{(\rho)} = i \left[\frac{I_{1}^{(\rho)}}{M_{\rho}^{2}} + I_{2M}^{(\rho)}\right] \left(p_{K^{0}} - p_{K^{-}}\right)_{\mu}, \nonumber\\
    	F_{\mu}^{(\omega)} = i \left[\frac{I_{1M}^{(\omega)}}{M_{\omega}^{2}} + I_{2M}^{(\omega)}\right] \left(p_{K^{0}} - p_{K^{-}}\right)_{\mu}, \nonumber\\
    	F_{\mu}^{(\phi)} = i \left[\frac{I_{1M}^{(\phi)}}{M_{\phi}^{2}} + I_{2M}^{(\phi)}\right] \left(p_{K^{0}} - p_{K^{-}}\right)_{\mu}.
    \end{eqnarray}
    
    The full amplitude taking into account the interaction in the final state takes the following form:
    \begin{eqnarray}
        M_{tot} & = &-2 \sqrt{2} G_{f} V_{ud} I_{11}^{KK} \left[T_{K}^{(c)} + \frac{C_{\rho}}{g_{\rho}} \frac{I_{11}^{KK\rho}}{I_{11}^{KK}} T_{K}^{(\rho)} \frac{q^2}{M_{\rho}^{2} - q^{2} - i\sqrt{q^{2}}\Gamma_{\rho}} \right. \nonumber\\
        && \left.+ \frac{C_{\rho'}}{g_{\rho}} \frac{I_{11}^{KK\rho'}}{I_{11}^{KK}} T_{K}^{(\rho')} \frac{q^2}{M_{\rho'}^{2} - q^{2} - i\sqrt{q^{2}}\Gamma_{\rho'}}\right] \left\{1 - 4 \left(I_{11}^{KK\rho}\right)^{2} \left(T_{K}^{(\rho)}\right)^{2} \left[\frac{I_{1}^{(\rho)}}{M_{\rho}^{2}} + I_{2M}^{(\rho)}\right]\right. \nonumber\\
        && \left. + 4 \left(I_{11}^{KK\omega}\right)^{2} \left(T_{K}^{(\omega)}\right)^{2} \left[\frac{I_{1}^{(\omega)}}{M_{\omega}^{2}} + I_{2M}^{(\omega)}\right] + 4 \left(I_{11}^{KK\phi}\right)^{2} \left(T_{K}^{(\phi)}\right)^{2} \left[\frac{I_{1}^{(\phi)}}{M_{\phi}^{2}} + I_{2M}^{(\phi)}\right]\right\} \nonumber\\
        && \times L_{\mu} \left(p_{K^{0}} -p_{K^{-}}\right)^{\mu},
    \end{eqnarray}
    where
    \begin{eqnarray}
        T_{K}^{(\omega)} & = & 1 - \frac{I_{11}^{K_{1}K\omega}I_{11}^{K_{1}K}}{I_{11}^{KK\omega}} \frac{\left(m_{s} + m_{u}\right)^{2}}{M_{K_{1A}}^{2}}, \nonumber\\
        T_{K}^{(\phi)} & = & 1 - \frac{I_{11}^{K_{1}K\phi}I_{11}^{K_{1}K}}{I_{11}^{KK\phi}} \frac{\left(m_{s} + m_{u}\right)^{2}}{M_{K_{1A}}^{2}}, \nonumber\\
		I_{2M}^{(\rho)} =
		\frac{-i}{(2\pi)^{4}}\int\frac{\theta(\Lambda_{M}^{2} + k^2)}{(M_{\rho}^{2} - k^2)(M_{K}^{2} - k^2)} \mathrm{d}^{4}k & = & \frac{1}{(4\pi)^{2}}\frac{1}{M_{\rho}^{2} - M_{K}^{2}}\left[M_{\rho}^{2}\ln\left(\frac{\Lambda_{M}^{2}}{M_{\rho}^{2}} + 1\right) - M_{K}^{2}\ln\left(\frac{\Lambda_{M}^{2}}{M_{K}^{2}} + 1\right)\right], \nonumber\\
		I_{2M}^{(\omega)} =
		\frac{-i}{(2\pi)^{4}}\int\frac{\theta(\Lambda_{M}^{2} + k^2)}{(M_{\omega}^{2} - k^2)(M_{K}^{2} - k^2)} \mathrm{d}^{4}k & = & \frac{1}{(4\pi)^{2}}\frac{1}{M_{\omega}^{2} - M_{K}^{2}}\left[M_{\omega}^{2}\ln\left(\frac{\Lambda_{M}^{2}}{M_{\omega}^{2}} + 1\right) - M_{K}^{2}\ln\left(\frac{\Lambda_{M}^{2}}{M_{K}^{2}} + 1\right)\right], \nonumber\\
		I_{2M}^{(\phi)} =
		\frac{-i}{(2\pi)^{4}}\int\frac{\theta(\Lambda_{M}^{2} + k^2)}{(M_{\phi}^{2} - k^2)(M_{K}^{2} - k^2)} \mathrm{d}^{4}k & = & \frac{1}{(4\pi)^{2}}\frac{1}{M_{\phi}^{2} - M_{K}^{2}}\left[M_{\phi}^{2}\ln\left(\frac{\Lambda_{M}^{2}}{M_{\phi}^{2}} + 1\right) - M_{K}^{2}\ln\left(\frac{\Lambda_{M}^{2}}{M_{K}^{2}} + 1\right)\right], \nonumber\\
		I_{1M}^{(\rho)} = \frac{-i}{(2\pi)^{4}}\int\frac{\theta(\Lambda_{M}^{2} + k^2)}{(M_{\rho}^{2} - k^2)} \mathrm{d}^{4}k & = & \frac{1}{(4\pi)^{2}} \left[\Lambda_{M}^{2} - M_{\rho}^{2}\ln\left(\frac{\Lambda_{M}^{2}}{M_{\rho}^{2}} + 1\right)\right], \nonumber\\
		I_{1M}^{(\omega)} = \frac{-i}{(2\pi)^{4}}\int\frac{\theta(\Lambda_{M}^{2} + k^2)}{(M_{\omega}^{2} - k^2)} \mathrm{d}^{4}k & = & \frac{1}{(4\pi)^{2}} \left[\Lambda_{M}^{2} - M_{\omega}^{2}\ln\left(\frac{\Lambda_{M}^{2}}{M_{\omega}^{2}} + 1\right)\right], \nonumber\\
		I_{1M}^{(\phi)} = \frac{-i}{(2\pi)^{4}}\int\frac{\theta(\Lambda_{M}^{2} + k^2)}{(M_{\phi}^{2} - k^2)} \mathrm{d}^{4}k & = & \frac{1}{(4\pi)^{2}} \left[\Lambda_{M}^{2} - M_{\phi}^{2}\ln\left(\frac{\Lambda_{M}^{2}}{M_{\phi}^{2}} + 1\right)\right].
	\end{eqnarray}
    
    The expression in the curly brackets includes the contributions of the meson triangles. The contributions of the triangles with the exchange of the $\rho$ and $\omega$ mesons approximately reduce each other and the triangle with the $\phi$ meson gives the main contribution.
    
    As a result of taking into account the interaction in the final state a new parameter, a cutoff parameter of the meson loop $\Lambda_{M}$, appears.
    
    In the previous section, the partial width of the decay $\tau \to K K \nu_{\tau}$ was calculated in the framework of the extended NJL model in satisfactory agreement with the experimental data within the model uncertainties. Taking into account the meson interaction in the final state and taking the value $\Lambda_{M} = 610$~MeV, one can obtain a precise agreement with the experimental data. This cutoff value is close to the one which was applied when describing the process $\tau \to \pi \pi \nu_{\tau}$ (740~MeV) and which was obtained using the related process $e^{+}e^{-} \to \pi^{+} \pi^{-}$ \cite{Volkov:2020dvz}. However, in that process, the inclusion of the meson interaction in the final state played an important role and gave a significant correction. In our case, this correction is smaller and not beyond the scope of the model uncertainties. The corrections of the same order could be obtained by variation of the intermediate radially excited meson decay width in the framework of the experimental errors. Also, it is interesting to note that when describing the related process $e^{+}e^{-} \to K^{+} K^{-}$ in the extended NJL model a satisfactory agreement with the experimental data was obtained without taking into account the meson interaction in the final state \cite{Volkov:2018cqp}.
    
\section{Conclusion}
    In our latest works, all possible tau decays into two pseudoscalar mesons have been described. The calculations have shown that taking into account the interaction in the final state plays an important role in such processes as $\tau \to \pi \pi \nu_{\tau}$ and $\tau \to K \pi \nu_{\tau}$ and its contribution is about 30\%. It should be noticed that in the pointed decays, in the vector channel the contribution of the ground state was taken into account while the influence of the intermediate radially excited vector mesons was insignificant. In the decays $\tau \to K \eta \nu_{\tau}$ and $\tau \to K K \nu_{\tau}$, with increasing the summary mass of the produced mesons, the role of the final state interaction decreases noticeably. In the decay $\tau \to K \eta \nu_{\tau}$, the contribution from the interaction in the final state has decreased to 15\%. In the decay $\tau \to K K \nu_{\tau}$, it has turned out to be about 6\% and is not beyond the scope of the model uncertainties. It is interesting to note that the role of the intermediate excited states in the vector channel has increased significantly in these processes. It is quite possible that this is related to the decreasing contribution of the final state interaction. As it was shown in section 3, changing the intermediate excited meson decay width influences the result stronger than taking into account the interaction in the final state. The nature of the mutual influence of the contributions of the intermediate radially excited states and meson interaction in the final state requires more careful study.

\section*{Acknowledgments}
    The authors are grateful to K. Nurlan for his interest in this work and useful discussions.

\end{document}